\documentclass[conference]{IEEEtran}
\IEEEoverridecommandlockouts
\usepackage{cite}
\usepackage{amsmath,amssymb,amsfonts}
\usepackage{graphicx}
\usepackage{subcaption}
\usepackage{textcomp}
\usepackage{xcolor}
\usepackage{algorithm}
\usepackage{algpseudocode}
\usepackage{epstopdf}
\usepackage{bm}

\def\BibTeX{{\rm B\kern-.05em{\sc i\kern-.025em b}\kern-.08em
	T\kern-.1667em\lower.7ex\hbox{E}\kern-.125emX}}

\begin{document}

\title{On the Capacity of Pixel Antenna based MIMO Communication}

\author{Shenrui Lin and Shuowen Zhang \\
	Department of Electrical and Electronic Engineering, The Hong Kong Polytechnic University \\
	E-mails: shenrui.lin@connect.polyu.hk, shuowen.zhang@polyu.edu.hk
}

\maketitle

\begin{abstract}
Pixel antenna is a promising technology to enhance the wireless communication data rate by adaptively reconfiguring each antenna's radiation pattern via a so-called antenna coding technique which controls the states of switches connected to multiple pixel ports. This paper studies a multiple-input multiple-output (MIMO) system where both the transmitter and the receiver are equipped with multiple pixel antennas. We aim to characterize the fundamental capacity limit of this MIMO system by jointly optimizing the transmit covariance matrix and the antenna coders at both the transmitter and the receiver. This problem is a mixed-integer non-linear program (MINLP) which is non-convex and particularly challenging to solve due to the binary-valued optimization variables corresponding to the antenna coders. We first propose an exhaustive search based method to obtain the optimal solution to this problem, which corresponds to the fundamental capacity limit. Then, we propose a branch-and-bound based iterative algorithm aiming to find a high-quality suboptimal solution with lower complexity than exhaustive search as the number of pixel ports becomes large. Finally, we devise an alternating optimization (AO) based algorithm with polynomial complexity. Numerical results show that our proposed algorithms achieve a flexible trade-off between performance and complexity. Moreover, equipping the transceivers with pixel antennas can enhance the achievable rate of MIMO communications.
\end{abstract}

\section{Introduction}
Multiple-input multiple-output (MIMO) has been one of the most important technologies in wireless communications. Over the past few years, various techniques have been developed to amplify the performance gain of MIMO communications, including massive MIMO \cite{lu2014}, hybrid analog-digital MIMO beamforming \cite{Yu}, and reconfigurable intelligent surface (RIS) aided MIMO \cite{Zhang}. However, the antennas considered in vast majority of the existing studies have fixed radiation patterns, which, together with the MIMO channel, need to be accommodated using beamforming and other signal processing techniques. As future communication systems strive for extremely high data rate performance, it is imperative to seek a paradigm shift in antenna technology to open up new degrees-of-freedom (DoFs) in terms of the radiation pattern.

Pixel antenna is a promising technology that offers the possibility of radiation pattern reconfiguration \cite{pringle2004}. The concept is based on discretizing a continuous radiation surface into small elements, referred to as pixels. Adjacent pixels can be connected through radio frequency (RF) switches \cite{grau2011}, \cite{rodrigo2014}, whose states collaboratively determine the actual radiation pattern of the antenna. By judiciously designing the states of the switches, the antennas' radiation patterns can be changed to enhance the overall effective MIMO channel. Due to the ``on'' and ``off'' states of each switch, such design is also termed as antenna coding \cite{shen2025,li2025}. Different from conventional antenna selection where each selected antenna can only be connected to one RF chain with only one possible radiation pattern, each pixel antenna with a large number of pixels and switches can be connected to one RF chain to create a large number of possible radiation patterns. Moreover, the overall effective MIMO channels with pixel antenna and antenna selection have fundamentally different mathematical expressions due to the intrinsically different ways the MIMO channel is changed. 

The potential of pixel antenna and antenna coding has been recently explored. \cite{shen2025} introduced an electromagnetic-based communication model for pixel antennas and optimized antenna coding to maximize the channel gain of single-input single-output (SISO) systems. A heuristic algorithm termed successive exhaustive boolean optimization (SEBO) \cite{shen2017} was employed which divides the binary optimization variables into multiple blocks. This concept was subsequently extended in \cite{li2025} to multi-user multiple-input single-output (MISO) systems, where antenna coding at the user side was jointly optimized with the transmitter's precoder to maximize the sum-rate via a similar SEBO method. These works showed the significant performance enhancement enabled by pixel antennas through the adaptive reconfiguration of radiation patterns.

However, the complexity of the SEBO method is exponential in the number of binary optimization variables allocated to each block. How to devise a scalable polynomial-complexity optimization algorithm for the binary antenna coders remains open for investigation. Moreover, the fundamental capacity limit of pixel antenna based MIMO communication is still unknown. In this paper, we aim to fill these research gaps by studying a pixel antenna based MIMO system where both the transmitter and the receiver are equipped with multiple pixel antennas. To characterize the fundamental capacity limit, we study the joint optimization problem of the transmit covariance matrix and the antenna coding matrices at the transmitter and the receiver to maximize the achievable rate. This problem is a non-convex mixed-integer non-linear program (MINLP) which is difficult to solve. We first propose an exhaustive search method to find the optimal solution, which corresponds to the capacity limit of pixel antenna based MIMO communication. Then, we propose two high-quality suboptimal solutions via branch-and-bound or element-wise alternating optimization (AO), where the latter has polynomial time complexity. It is shown via numerical results that the proposed algorithms achieve a flexible trade-off between rate and complexity, and outperform various benchmark schemes in data rate. Moreover, pixel antenna based MIMO communication has a significantly higher capacity compared to conventional MIMO communication even with a small number of switches or pixel ports.

\section{System Model}
In this section, we present the system model for a point-to-point MIMO communication system where both the transmitter and the receiver are equipped with multiple pixel antennas. We will first present the model for each pixel antenna, and then present the overall MIMO communication system.
\subsection{Modeling of Each Pixel Antenna}
Each pixel antenna can be modeled by using microwave multi-port network theory \cite{perruisseau2010}, \cite{yousefbeiki2012}. Specifically, a pixel antenna with $S$ RF switches is modeled as an ($S+1$)-port circuit network \cite{shen2025}. This network consists of one antenna port for signal feeding and $S$ pixel ports that represent the connections between adjacent pixels. The overall \emph{radiation pattern} of the pixel antenna is the superposition of the radiation patterns excited by the current at the antenna port and the currents at the $S$ pixel ports. To obtain the radiation pattern, we leverage the following model to derive these current values.

Let $v_{\mathrm{A}}$ and $\bm{v}_{\mathrm{P}}=[v_{{\mathrm{P}},1},...,v_{{\mathrm{P}},S}]^T\in\mathbb{C}^{S \times 1}$ denote the voltage at the antenna port and voltages at the pixel ports, respectively. Let $i_{\mathrm{A}}$ and $\bm{i}_{\mathrm{P}}=[i_{{\mathrm{P}},1},...,i_{{\mathrm{P}},S}]^T\in\mathbb{C}^{S \times 1}$ denote the current at the antenna port and currents at the pixel ports, respectively. The voltages are related to the currents via an impedance matrix $\bm{Z}\in \mathbb{C}^{(S+1)\times (S+1)}$ as shown below:
\begin{equation}\label{Z}
	\begin{bmatrix} v_{\mathrm{A}} \\ \bm{v}_{\mathrm{P}} \end{bmatrix}
	= \bm{Z}
	\begin{bmatrix} i_{\mathrm{A}} \\ \bm{i}_{\mathrm{P}} \end{bmatrix}=
	\begin{bmatrix} z_{\mathrm{AA}} & \bm{z}_{\mathrm{AP}}^T \\ \bm{z}_{\mathrm{PA}} & \bm{Z}_{\mathrm{PP}} \end{bmatrix}
	\begin{bmatrix} i_{\mathrm{A}} \\ \bm{i}_{\mathrm{P}} \end{bmatrix},
\end{equation}
where $z_{{\mathrm{AA}}} \in \mathbb{C}$ denotes the self-impedance of the antenna port; $\bm{Z}_{{\mathrm{PP}}} \in \mathbb{C}^{S \times S}$ denotes the impedance matrix for the $S$ pixel ports; $\bm{z}_{{\mathrm{AP}}}^T \in \mathbb{C}^{1 \times S}$ and $\bm{z}_{{\mathrm{PA}}} \in \mathbb{C}^{S \times 1}$ denote the trans-impedance matrices. It follows that $\bm{v}_{\mathrm{P}}=\bm{z}_{\mathrm{PA}}i_{\mathrm{A}}+\bm{Z}_{\mathrm{PP}}\bm{i}_{\mathrm{P}}$.

\begin{figure}[t]
	\centering
	\includegraphics[width=1\columnwidth]{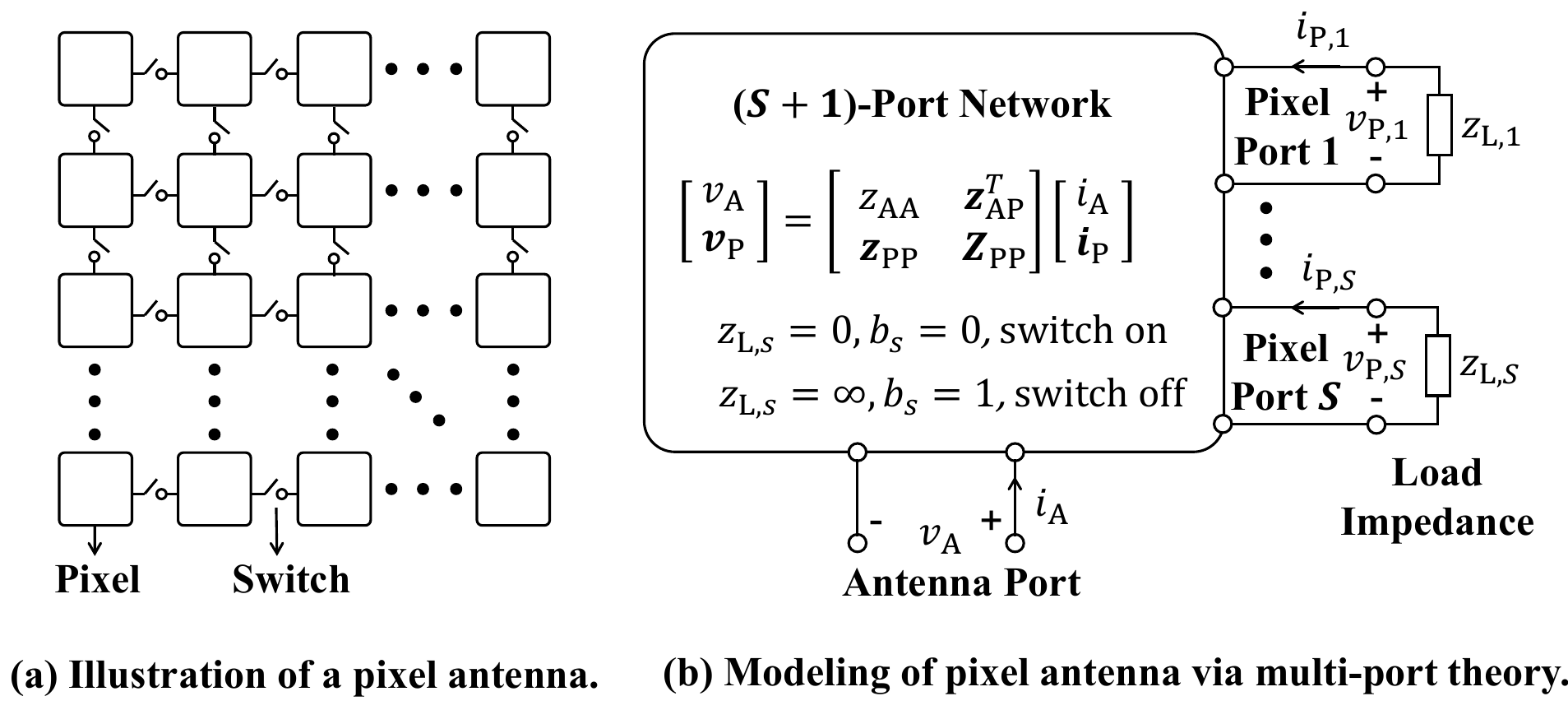}
	\vspace{-3mm}
	\caption{Illustration and modeling of pixel antenna.}\label{fig_system}
	\vspace{-6mm}
\end{figure}

Moreover, the switch connected to each $s$-th pixel port has an ``on'' state and an ``off'' state denoted by a binary variable $b_s=0$ or $b_s=1$, respectively. This corresponds to a load impedance circuit in the open circuit mode with impedance $z_{\mathrm{L},s}=Z_{\mathrm{on}}=0$ or short circuit mode with impedance $z_{\mathrm{L},s}=Z_{\mathrm{off}}\rightarrow\infty$, respectively, as illustrated in Fig. \ref{fig_system}.\footnote{$Z_{\mathrm{off}}$ can be set as a large finite value (e.g., $10^6$) for numerical tractability.} For ease of exposition, $z_{\mathrm{L},s}$ can be further expressed as
\begin{equation}
	z_{{\mathrm{L}},s} = (1 - b_s)  Z_{{\mathrm{on}}} + b_s  Z_{{\mathrm{off}}},\ s=1,...,S.
\end{equation}
Denote $\bm{b}=[b_1,...,b_S]^T$ as the collection of states of the switches connected to the $S$ pixels, which is also termed as the \emph{antenna coder} for the pixel antenna. The collection of load impedances of the $S$ pixel ports can be thus represented by 
\begin{align}
   	\bm{Z}_{\mathrm{L}}(\bm{b}) = &\mathrm{diag}\{z_{{\mathrm{L}},1},...,z_{{\mathrm{L}},S}\}\nonumber\\
   	=&Z_{{\mathrm{on}}}\bm{I}_S+(Z_{{\mathrm{off}}}-Z_{{\mathrm{on}}})\mathrm{diag}(\bm{b}).
\end{align}
Based on Fig. \ref{fig_system} and (\ref{Z}), we further have
\begin{align}
	\bm{v}_{\mathrm{P}}=-\bm{Z}_{\mathrm{L}}(\bm{b})\bm{i}_{\mathrm{P}}=\bm{z}_{\mathrm{PA}}i_{\mathrm{A}}+\bm{Z}_{\mathrm{PP}}\bm{i}_{\mathrm{P}}.
\end{align}
Thus, the currents at the pixel ports can be derived as 
\begin{align}
   	\bm{i}_{\mathrm{P}}=-(\bm{Z}_{\mathrm{L}}(\bm{b})+\bm{Z}_{\mathrm{PP}})^{-1}\bm{z}_{\mathrm{PA}}i_{\mathrm{A}}.
\end{align}

We consider a beamspace-based radiation model which discretizes the $2\pi$ spatial angle range into $K\geq 1$ sampled spatial angles \cite{shen2025}. Let $\bm{e}_{\mathrm{A}}\in \mathbb{C}^{2K\times 1}$ and $\bm{e}_{\mathrm{P},s}\in \mathbb{C}^{2K\times 1}$ denote the radiation pattern of the antenna port or each $s$-th pixel port over the $K$ angles and two polarizations excited by a unit current when all the other ports are open-circuited, respectively. Then, the overall radiation pattern is the weighted summation of $\bm{e}_{\mathrm{A}}$ and $\bm{e}_{\mathrm{P},s}$'s with the weight being the current at the corresponding port, which is given by
\begin{align}\label{radiation}
		&\bm{e}(\bm{b}) = \bm{e}_{\mathrm{A}} i_{\mathrm{A}} + \sum_{s=1}^{S} \bm{e}_{{\mathrm{P}},s} i_{{\mathrm{P}},s}(\bm{b}) \\
		= &\left(\bm{e}_{\mathrm{A}} - [\bm{e}_{{\mathrm{P}},1},...,\bm{e}_{{\mathrm{P}},S}](\bm{Z}_{{\mathrm{PP}}} + \bm{Z}_{\mathrm{L}}(\bm{b}))^{-1} \bm{z}_{{\mathrm{PA}}}\right) i_{\mathrm{A}}.
\end{align}
The normalized (unit-power) radiation pattern is thus given by $\bar{\bm{e}}(\bm{b})=\frac{\bm{e}(\bm{b})}{\|\bm{e}(\bm{b})\|}$. Note that in the radiation pattern, $\bm{e}_{\mathrm{A}}$, $[\bm{e}_{{\mathrm{P}},1},...,\bm{e}_{{\mathrm{P}},S}]$, $\bm{Z}_{{\mathrm{PP}}}$, and $\bm{z}_{{\mathrm{PA}}}$ are constants determined by the given properties of the pixel antenna, while the only design variable is the antenna coder $\bm{b}$. By reconfiguring the antenna coder, different radiation patterns can be achieved.

Note that designing the antenna coder $\bm{b}$ is highly non-trivial due to the complex relationship between $\bar{\bm{e}}(\bm{b})$ and $\bm{b}$. Specifically, the design of $\bm{b}$ needs to consider all the $2K$ possible angle-polarization pairs and strike a desired balance among them in the radiation pattern. Moreover, the optimal design of $\bm{b}$ even for maximizing the power of a particular angle-polarization pair is not straightforward due to the lack of monotonicity. Thus, advanced optimization methods need to be developed for $\bm{b}$ based on the specific system of interest.

\subsection{Modeling of Pixel Antenna based MIMO Communication}
Based on the above model for each pixel antenna, we establish the model for pixel antenna based MIMO communication. Consider a MIMO system with $N_{\mathrm{T}}\geq 1$ pixel antennas at the transmitter and $N_{\mathrm{R}}\geq 1$ pixel antennas at the receiver. Each pixel antenna has $S$ pixel ports. Let $\bm{B}_{\mathrm{T}} = [\bm{b}_{{\mathrm{T}},1},..., \bm{b}_{{\mathrm{T}},N_{\mathrm{T}}}] \in \{0,1\}^{S \times N_{\mathrm{T}}}$ and $\bm{B}_{\mathrm{R}} = [\bm{b}_{{\mathrm{R}},1},...,\bm{b}_{{\mathrm{R}},N_{\mathrm{R}}}] \in \{0,1\}^{S \times N_{\mathrm{R}}}$ denote the antenna coder matrices for the transmitter and the receiver, respectively. The corresponding radiation pattern matrices $\bm{E}_{\mathrm{T}}(\bm{B}_{\mathrm{T}})$ and $\bm{E}_{\mathrm{R}}(\bm{B}_{\mathrm{R}})$ are formed by concatenating the normalized radiation patterns of each antenna, i.e., $\bm{E}_{\mathrm{T}}(\bm{B}_{\mathrm{T}}) = \left[\bar{\bm{e}}(\bm{b}_{{\mathrm{T}},1}),..., \bar{\bm{e}}(\bm{b}_{{\mathrm{T}},N_{\mathrm{T}}})\right] \in \mathbb{C}^{2K \times N_{\mathrm{T}}}$ and $\bm{E}_{\mathrm{R}}(\bm{B}_{\mathrm{R}}) = \left[\bar{\bm{e}}(\bm{b}_{{\mathrm{R}},1}),..., \bar{\bm{e}}(\bm{b}_{{\mathrm{R}},N_{\mathrm{R}}})\right] \in \mathbb{C}^{2K \times N_{\mathrm{R}}}$.

Let $\bm{H}_{\mathrm{V}} \in \mathbb{C}^{2K \times 2K}$ denote the ``virtual'' channel matrix which characterizes the physical propagation environment between the transmitter and the receiver over the $K$ sampled spatial angles and two polarizations. The overall effective MIMO channel matrix $\bm{H} \in \mathbb{C}^{N_{\mathrm{R}} \times N_{\mathrm{T}}}$ is determined by the antenna radiation patterns at the transmitter and the receiver as well as the physical propagation environment as
\begin{align}
	\bm{H}(\bm{B}_{\mathrm{T}}, \bm{B}_{\mathrm{R}}) = \bm{E}_{\mathrm{R}}^H(\bm{B}_{\mathrm{R}}) \bm{H}_{\mathrm{V}} \bm{E}_{\mathrm{T}}(\bm{B}_{\mathrm{T}}).
\end{align}  
	
Let $\bm{x}\sim \mathcal{CN}(\bm{0},\bm{Q})$ denote the transmitted signal vector where $\bm{Q} \triangleq \mathbb{E}[\bm{x}\bm{x}^H] \in \mathbb{C}^{N_{\mathrm{T}} \times N_{\mathrm{T}}}$ denotes the transmit covariance matrix, with $\bm{Q} \succeq \bm{0}$. We consider an average sum power constraint at the transmitter given by $\mathbb{E}[\|\bm{x}\|^2] \le P$, which is equivalent to $\mathrm{tr}(\bm{Q}) \le P$. The received signal vector denoted as $\bm{y} \in \mathbb{C}^{N_{\mathrm{R}} \times 1}$ is given by $\bm{y} = \bm{H}(\bm{B}_{\mathrm{T}}, \bm{B}_{\mathrm{R}}) \bm{x} + \bm{z} = \bm{E}_{\mathrm{R}}^H(\bm{B}_{\mathrm{R}}) \bm{H}_{\mathrm{V}} \bm{E}_{\mathrm{T}}(\bm{B}_{\mathrm{T}}) \bm{x} + \bm{z}$, where $\bm{z} \sim \mathcal{CN}(\bm{0}, \sigma^2 \bm{I}_{N_{\mathrm{R}}})$ denotes the circularly symmetric complex Gaussian (CSCG) noise vector at the receiver, with $\sigma^2$ denoting the average noise power. To reveal the fundamental capacity limit, we assume that $\bm{H}_{\mathrm{V}}$ is known at both the transmitter and the receiver. 

With given $\bm{B}_{\mathrm{T}}$, $\bm{B}_{\mathrm{R}}$, and $\bm{Q}$, the achievable rate is given by
\begin{align}
	&R(\bm{B}_{\mathrm{T}}, \bm{B}_{\mathrm{R}},\bm{Q})\nonumber\\
	=&\log_2 \det\biggl(\bm{I}_{N_{\mathrm{R}}} + \frac{1}{\sigma^2} \bm{H}(\bm{B}_{\mathrm{T}}, \bm{B}_{\mathrm{R}})	\bm{Q}\bm{H}^H(\bm{B}_{\mathrm{T}}, \bm{B}_{\mathrm{R}})\biggr).
\end{align}

Therefore, to characterize the capacity limit of pixel antenna based MIMO communication, $\bm{B}_{\mathrm{T}}$, $\bm{B}_{\mathrm{R}}$, and $\bm{Q}$ need to be jointly optimized to maximize $R(\bm{B}_{\mathrm{T}}, \bm{B}_{\mathrm{R}},\bm{Q})$. Note that this is a particularly challenging task for MIMO systems, since multiple data streams generally need to be sent through the channel for spatial multiplexing, while the power of each channel is determined by the overall effective MIMO channel $\bm{H}(\bm{B}_{\mathrm{T}}, \bm{B}_{\mathrm{R}})$. How to optimize $\bm{B}_{\mathrm{T}}$ and $\bm{B}_{\mathrm{R}}$ to create the optimal spatial sub-channels is a difficult problem.

\section{Problem Formulation}
To characterize the capacity of pixel antenna based MIMO communication, we formulate the following problem to jointly optimize the transmit covariance matrix $\bm{Q}$ as well as the binary transmit and receive antenna coders in $\bm{B}_{\mathrm{T}}$ and $\bm{B}_{\mathrm{R}}$:
\begin{align}
	(\text{P1})\quad &\nonumber \\
	\max_{\bm{B}_{\mathrm{T}}, \bm{B}_{\mathrm{R}}, \bm{Q}} & \log_2 \det\biggl(\bm{I}_{N_{\mathrm{R}}} + \frac{1}{\sigma^2} \bm{H}(\bm{B}_{\mathrm{T}}, \bm{B}_{\mathrm{R}}) \bm{Q}\bm{H}^H(\bm{B}_{\mathrm{T}}, \bm{B}_{\mathrm{R}})\biggr) \\
	\text{s.t.} \quad & [\bm{B}_{\mathrm{T}}]_{i,j} \in \{0, 1\}, \ i=1,\dots,S, \ j=1,\dots,N_{\mathrm{T}} \!\!\\
	& [\bm{B}_{\mathrm{R}}]_{i,j} \in \{0, 1\}, \ i=1,\dots,S, \ j=1,\dots,N_{\mathrm{R}} \!\!\\
	& \operatorname{tr}(\bm{Q}) \le P \\
	& \bm{Q} \succeq \bm{0}.
\end{align}

(P1) is a non-convex optimization problem due to the discrete nature of the variables in $\bm{B}_{\mathrm{T}}$ and $\bm{B}_{\mathrm{R}}$ as well as the non-concave objective function. Moreover, the coupling among $\bm{Q}$, $\bm{B}_{\mathrm{T}}$, and $\bm{B}_{\mathrm{R}}$ makes (P1) more challenging to solve. The mixture of discrete variables in $\bm{B}_{\mathrm{T}}$ and $\bm{B}_{\mathrm{R}}$ and continuous variables in $\bm{Q}$ further makes (P1) a challenging MINLP. In the following, we present both the optimal solution to (P1) and lower-complexity alternative solutions.
\section{Optimal Solution to (P1)}
Note that with given $\bm{B}_{\mathrm{T}}$, $\bm{B}_{\mathrm{R}}$ and consequently the effective MIMO channel $\bm{H}(\bm{B}_{\mathrm{T}}, \bm{B}_{\mathrm{R}})$, (P1) reduces to the following optimization problem on the transmit covariance matrix $\bm{Q}$:
\begin{align}
	(\text{P1-Q})\nonumber \\
	\max_{\bm{Q}}\ & \log_2 \det\biggl(\bm{I}_{N_{\mathrm{R}}} + \frac{1}{\sigma^2} \bm{H}(\bm{B}_{\mathrm{T}}, \bm{B}_{\mathrm{R}}) \bm{Q}\bm{H}^H(\bm{B}_{\mathrm{T}}, \bm{B}_{\mathrm{R}})\biggr) \\
	\text{s.t.} \ & \operatorname{tr}(\bm{Q}) \le P \\
	& \bm{Q} \succeq \bm{0}.
\end{align}

(P1-Q) is a convex optimization problem, for which the optimal solution is given by eigenmode transmission \cite{goldsmith2005}. Specifically, let $D = \mathrm{rank}(\bm{H}(\bm{B}_{\mathrm{T}}, \bm{B}_{\mathrm{R}})) \le \min(N_{\mathrm{T}}, N_{\mathrm{R}})$ denote the rank of the effective MIMO channel $\bm{H}(\bm{B}_{\mathrm{T}}, \bm{B}_{\mathrm{R}})$. Denote $\bm{H}(\bm{B}_{\mathrm{T}}, \bm{B}_{\mathrm{R}}) = \bm{U} \bm{\Lambda} \bm{V}^H$ as the truncated singular-value decomposition (SVD) of $\bm{H}(\bm{B}_{\mathrm{T}}, \bm{B}_{\mathrm{R}})$, where $\bm{U}\in \mathbb{C}^{N_{\mathrm{R}}\times D}$, $\bm{V}\in \mathbb{C}^{N_{\mathrm{T}}\times D}$, $\bm{\Lambda}=\mathrm{diag}\{\lambda_1,...,\lambda_D\}$, with $\bm{U}^H\bm{U}=\bm{I}_{D}$ and $\bm{V}^H\bm{V}=\bm{I}_{D}$. The optimal transmission strategy (i.e., the optimal solution to (P1-Q)) decomposes the effective MIMO channel into $D$ parallel spatial sub-channels via
\begin{align}\label{Q}
	\bm{Q}^\star(\bm{B}_{\mathrm{T}}, \bm{B}_{\mathrm{R}}) = \bm{V} \bm{\Sigma} \bm{V}^H,
\end{align}
where $\bm{\Sigma}=\mathrm{diag}\{p_1,...,p_D\}$, with $p_i = \max\{\nu-\sigma^2/\lambda_i^2,0\}$ denoting the power allocated to the $i$-th sub-channel and $\sum_{i=1}^D p_i=P$.

Based on this, the optimal solution to (P1) can be obtained by solving (P1-Q) for all feasible solutions of $\bm{B}_{\mathrm{T}}, \bm{B}_{\mathrm{R}}$, and then selecting the one with the maximum achievable rate. Note that there are $2^{SN_{\mathrm{T}}}$ and $2^{SN_{\mathrm{R}}}$ feasible solutions for $\bm{B}_{\mathrm{T}}$ and $\bm{B}_{\mathrm{R}}$, respectively, which correspond to $2^{S(N_{\mathrm{T}}+N_{\mathrm{R}})}$ feasible solutions of $\bm{B}_{\mathrm{T}},\bm{B}_{\mathrm{R}}$. By further noting that solving (P1-Q) requires a worst-case complexity of $\mathcal{O}((N_{\mathrm{T}} + N_{\mathrm{R}})(S^3 + 2KS) + 4K^2 \min(N_{\mathrm{R}}, N_{\mathrm{T}}) + 2KN_{\mathrm{R}}N_{\mathrm{T}} + N_{\mathrm{T}} N_{\mathrm{R}} \min(N_{\mathrm{T}}, N_{\mathrm{R}}))$, the overall complexity for obtaining the optimal solution to (P1) via exhaustive search is given by $\mathcal{O}(2^{S(N_{\mathrm{T}}+N_{\mathrm{R}})}((N_{\mathrm{T}} + N_{\mathrm{R}})(S^3 + 2KS) + 4K^2 \min(N_{\mathrm{R}}, N_{\mathrm{T}}) + 2KN_{\mathrm{R}}N_{\mathrm{T}} + N_{\mathrm{T}} N_{\mathrm{R}} \min(N_{\mathrm{T}}, N_{\mathrm{R}})))$. Note that the obtained solution corresponds to the fundamental capacity limit of pixel antenna based MIMO communication.

\section{Branch-and-Bound based Solution to (P1)}
Although the exhaustive search method is capable of finding the optimal solution to (P1), the complexity of it may be prohibitive when the problem dimension (e.g., $S$) becomes large, which limits its scalability. In this section, we present a branch-and-bound based solution to (P1) for the purpose of reducing the complexity especially when $S$ is large.

Specifically, we propose to iteratively optimize the continuous variables in $\bm{Q}$ and the discrete variables in $\bm{B}_{\mathrm{T}}, \bm{B}_{\mathrm{R}}$ until convergence. The sub-problem for optimizing $\bm{Q}$ is given in (P1-Q), for which the optimal solution is given by (\ref{Q}). On the other hand, the sub-problem of optimizing $\bm{B}_{\mathrm{T}}, \bm{B}_{\mathrm{R}}$ with given $\bm{Q}$ is formulated as
\begin{align}
	(\text{P1-B}) \nonumber\\
	\max_{\bm{B}_{\mathrm{T}}, \bm{B}_{\mathrm{R}}}  & \log_2 \det\left(\bm{I}_{N_{\mathrm{R}}} + \frac{1}{\sigma^2} \bm{H}(\bm{B}_{\mathrm{T}}, \bm{B}_{\mathrm{R}}) \bm{Q} \bm{H}^H(\bm{B}_{\mathrm{T}}, \bm{B}_{\mathrm{R}})\right) \\
	\text{s.t.} \quad & [\bm{B}_{\mathrm{T}}]_{i,j} \in \{0, 1\},\  i=1,\dots,S,\ j=1,\dots,N_{\mathrm{T}} \\
	\phantom{\text{s.t.} \quad} & [\bm{B}_{\mathrm{R}}]_{i,j} \in \{0, 1\},\  i=1,\dots,S, \ j=1,\dots,N_{\mathrm{R}}.
\end{align}

We aim to obtain an optimal solution to (P1-B) via the branch-and-bound method. Specifically, we vectorize the transmit antenna coder matrix $\bm{B}_{\mathrm{T}} \in \{0, 1\}^{S \times N_{\mathrm{T}}}$ and the receive antenna coder matrix $\bm{B}_{\mathrm{R}} \in \{0, 1\}^{S \times N_{\mathrm{R}}}$ and concatenate them into a single binary decision vector denoted by $\bm{b}_{\text{all}}\in \{0,1\}^{S(N_{\mathrm{T}} + N_{\mathrm{R}})\times 1}$, which is the new optimization variable. The branch-and-bound method systematically explores the solution space of $\bm{b}_{\text{all}}$ by constructing a search tree and pruning branches that are proven not to contain a better solution than the ones already found \cite{fischetti2010}.

Specifically, the first step is initialization which critically determines the efficiency of the branch-and-bound algorithm. Although an initial lower bound of the achievable rate can be constructed as $0$, it may limit the effectiveness of pruning in the early stage of the search. To establish a strong incumbent solution, we employ a warm-start strategy. We first run a Diving Heuristic \cite{fischetti2010}, a multi-stage process combining greedy search with local refinement, to find a high-quality initial lower bound $R_{\text{best}}$, which is then used as the first global pruning threshold for the branch-and-bound search.

We perform bounding by calculating a tight upper bound of the maximum achievable rate. However, the non-convex and binary nature of the problem makes standard relaxation-based approaches computationally intractable. We then develop a hierarchical bounding strategy. When the number of unfixed variables at a node in the tree is small, we perform an exhaustive search to find the exact local optimum and use it as the updated global lower bound. For the general case, we compute an upper bound based on matrix norm theory. First, we estimate the sub-problem's maximum effective channel power as $W_{\text{adj}}= \gamma(f)N_{\mathrm{R}} N_{\mathrm{T}}\|\bm{H}_\mathrm{V}\|_F^2$, where $f$ is the ratio between the number of free variables and the total number of variables $S(N_{\mathrm{T}}+N_{\mathrm{R}})$, $\gamma(f)=c_1+c_2f$ with $c_1$ and $c_2$ being positive constants. We then characterize an ideal channel with $T_{\text{star}} = \min(N_{\mathrm{T}}, N_{\mathrm{R}})$ equal eigenvalues $\lambda_{\text{ideal}} = W_{\text{adj}}/T_{\text{star}}$, since this will maximize the channel capacity given a total channel power. An upper bound can be then obtained by further optimizing $\bm{Q}$ for this given channel.

Regarding branching, we define $n$ as the index for the binary vector $\bm{b}_{\text{all}}$. To select the next index $n$ to branch on, our algorithm learns from its progress by maintaining pseudocosts. Specifically, $\Psi_{n,0}$ and $\Psi_{n,1}$ represent the learned average degradation in the objective upper bound observed when the variable $\bm{b}_{\text{all}}[n]$ has been historically branched on to 0 or 1, respectively. They are updated each time a child node's bound is calculated. At any given node, the index $n^\star $ selected is chosen from all indices $n$ that are still free (unfixed) at that node, and is the one predicted to be most influential, given by $n^\star = \arg \max_{n} (\Psi_{n,0} + \Psi_{n,1})$.
	
Regarding pruning, we can perform standard pruning by removing a node if its upper bound is no larger than the current best rate. We may also perform more aggressive pruning to improve the efficiency. For example, primal heuristics can be periodically executed which takes the current best solution and applies a rapid local search to actively seek a new improved feasible solution. Moreover, dynamic pruning can be triggered if a primal heuristic successfully finds a superior solution, where the global lower bound can be immediately updated, and the algorithm iterates through the entire queue of unevaluated nodes and prunes any node in the queue whose upper bound is no larger than the newly updated global lower bound.

With standard pruning, the proposed branch-and-bound algorithm is guaranteed to find the optimal solution to (P1-B). The worst-case complexity of this algorithm for (P1-B) can be shown to be $\mathcal{O}(I_d S^5(N_{\mathrm{T}}+N_{\mathrm{R}})^3 + I_d S^2(N_{\mathrm{T}}+N_{\mathrm{R}})^3 K^2 + I_d S^2(N_{\mathrm{T}}+N_{\mathrm{R}})^2 N_{\mathrm{R}}N_{\mathrm{T}}(N_{\mathrm{T}}+N_{\mathrm{R}}) + \beta N_b S^4(N_{\mathrm{T}}+N_{\mathrm{R}})^2 + \beta N_b S(N_{\mathrm{T}}+N_{\mathrm{R}})^2 K^2 + \beta N_b S(N_{\mathrm{T}}+N_{\mathrm{R}}) N_{\mathrm{T}}N_{\mathrm{R}}(N_{\mathrm{T}}+N_{\mathrm{R}}))$ with $I_d$ denoting the number of Diving iterations, $N_b$ denoting the number of nodes visited, and $\beta$ representing the node reduction factor from warm-start initialization. Although the worst-case complexity is generally still exponential in $S$ since the worst-case number of nodes visited, $N_b$, is exponential in $S$, this algorithm is anticipated to incur lower complexity compared to the exhaustive search method for solving (P1-B).
 
To summarize, (P1-Q) and (P1-B) can be iteratively solved via (\ref{Q}) or the branch-and-bound algorithm, respectively, until convergence. Let $I_b$ denote the number of iterations. The overall complexity is given by $\mathcal{O}(I_b(I_d S^5(N_{\mathrm{T}}+N_{\mathrm{R}})^3 + I_d S^2(N_{\mathrm{T}}+N_{\mathrm{R}})^3 K^2 + I_d S^2(N_{\mathrm{T}}+N_{\mathrm{R}})^2 N_{\mathrm{R}}N_{\mathrm{T}}(N_{\mathrm{T}}+N_{\mathrm{R}}) + \beta N_b S^4(N_{\mathrm{T}}+N_{\mathrm{R}})^2 + \beta N_b S(N_{\mathrm{T}}+N_{\mathrm{R}})^2 K^2 + \beta N_b S(N_{\mathrm{T}}+N_{\mathrm{R}}) N_{\mathrm{T}}N_{\mathrm{R}}(N_{\mathrm{T}}+N_{\mathrm{R}}))((N_{\mathrm{T}} + N_{\mathrm{R}})(S^3 + 2KS) + 4K^2 \min(N_{\mathrm{R}}, N_{\mathrm{T}}) + 2KN_{\mathrm{R}}N_{\mathrm{T}} + N_{\mathrm{T}} N_{\mathrm{R}} \min(N_{\mathrm{T}}, N_{\mathrm{R}})))$. 
\section{Element-wise AO based Solution to (P1)}
Finally, we propose a low-complexity algorithm for (P1) via element-wise AO. Specifically, we aim to iteratively optimize $\bm{Q}$ or an entry in $\bm{B}_{\mathrm{T}}$ or $\bm{B}_{\mathrm{R}}$. The sub-problem for optimizing $\bm{Q}$ is given in (P1-Q), for which the optimal solution is given by (\ref{Q}). The optimal solution to each $[\bm{B}_{\mathrm{T}}]_{i,j}$ or $[\bm{B}_{\mathrm{R}}]_{i,j}$ can be simply obtained by comparing the achievable rates $R(\bm{B}_{\mathrm{T}}, \bm{B}_{\mathrm{R}},\bm{Q})$ with switch states ``on'' and ``off''. Namely, 
\begin{align}
	[\bm{B}_{\mathrm{T}}]_{i,j}^\star = \arg \max_{b \in \{0,1\}} R(\bm{B}_{\mathrm{T}}|_{[\bm{B}_{\mathrm{T}}]_{i,j} = b}, \bm{B}_{\mathrm{R}},\bm{Q}),\label{B_T}\\
	[\bm{B}_{\mathrm{R}}]_{i,j}^\star = \arg \max_{b \in \{0,1\}} R(\bm{B}_{\mathrm{T}},\bm{B}_{\mathrm{R}}|_{[\bm{B}_{\mathrm{R}}]_{i,j} = b},\bm{Q}).\label{B_R}
\end{align}
Note that since the optimal solution to each sub-problem is obtained in each iteration, the element-wise AO based algorithm is guaranteed to monotonically converge.

Since the quality of the AO based solution is critically determined by the initial point, we propose to randomly generate $L\geq 1$ realizations of $\bm{B}_{\mathrm{T}}$, $\bm{B}_{\mathrm{R}}$. Based on each initial point, we iteratively find the optimal solution to $\bm{Q}$, $[\bm{B}_{\mathrm{T}}]_{i,j}$ or $[\bm{B}_{\mathrm{R}}]_{i,j}$ via (\ref{Q}), (\ref{B_T}), or (\ref{B_R}), until convergence. Finally, we select the best converged solution among the $L$ ones.

Let $I$ denote the worst-case number of iterations for the proposed AO based algorithm. The worst-case complexity for this algorithm can be shown to be $\mathcal{O}(LI(S^4(N_{\mathrm{T}}+N_{\mathrm{R}})^2 + K S^2(N_{\mathrm{T}}+N_{\mathrm{R}})^2 + K^2 S(N_{\mathrm{T}}+N_{\mathrm{R}})^2 + S(N_{\mathrm{T}}+N_{\mathrm{R}})\min\{N_{\mathrm{T}},N_{\mathrm{R}}\}^2 \max\{N_{\mathrm{T}},N_{\mathrm{R}}\}))$.

\section{Numerical Results}
\begin{figure}[t]
	\centering
	\includegraphics[width=3cm]{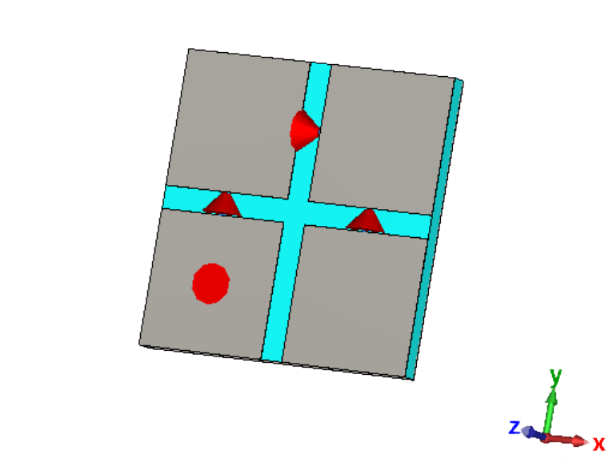}
			\vspace{-2mm}
	\caption{Illustration of a pixel antenna model with $S=3$ in CST Studio Suite.}
	\label{fig_pixel}
		\vspace{-7mm}
\end{figure}

In this section, we evaluate the performance of our proposed designs for MIMO communication with pixel antennas. We consider a rich-scattering propagation environment with $[\bm{H}_\mathrm{V}]_{i,j}\sim \mathcal{CN}(0, \beta)$, where $\beta$ denotes the distance-dependent channel power modeled by $\beta = \beta_0(d/d_0)^{-\overline{\alpha}}$, with $\beta_0 = -30 \text{ dB}$ denoting the channel power at the reference distance $d_0=1 \text{ m}$; $d=600$ m denoting the transmitter-receiver distance; and $\overline{\alpha}=3.5$ denoting the path loss exponent. We set $\sigma^2 = -90~\text{dBm}$. The (receiver) signal-to-noise ratio (SNR) is defined as $\frac{P\beta}{\sigma^2}$. We further set $N_{\mathrm{T}}=N_{\mathrm{R}}=2$ and $S=3$ with a configuration shown in Fig. \ref{fig_pixel} unless specified otherwise, for which the impedance matrices, fundamental radiation patterns, and other properties of the pixel antennas are simulated by CST Studio Suite at 2.4 GHz. All results are averaged over 100 independent realizations of $\bm{H}_{\mathrm{V}}$.

Besides our proposed exhaustive search method, branch-and-bound method (denoted by BnB), and the element-wise AO method (denoted by AO), we consider the following benchmark schemes for comparison:
\begin{itemize}
	\item {\textbf{SEBO}} \cite{shen2025}: This method divides the binary variables in $\bm{B}_{\mathrm{T}}$ and $\bm{B}_{\mathrm{R}}$ into multiple blocks each with $J=3$ binary variables. Each block of variables or $\bm{Q}$ is iteratively optimized via exhaustive search or (\ref{Q}), until convergence is reached. A random bit-flipping procedure is applied after convergence to seek further performance improvement.
	\item {\textbf{Conventional MIMO}}: This scheme considers a conventional MIMO system without pixel antennas, where the radiation pattern is fixed. The transmit covariance matrix is optimized in a similar manner as in (\ref{Q}).
	\item {\textbf{Best-Single Off}}: This scheme only selects one pixel port to be in the ``off'' state, while all the other ports are in the ``on'' state. The port selection is optimized via one-dimensional exhaustive search.
	\item {\textbf{Best-Single On}}: This scheme only selects one pixel port to be in the ``on'' state, while all the other ports are in the ``off'' state. The port selection is optimized via one-dimensional exhaustive search.
	\item {\textbf{Random-Single Off}}: This scheme randomly selects one pixel port to be in the ``off'' state, while all the other ports are in the ``on'' state.
	\item {\textbf{Random-Single On}}: This scheme randomly selects one pixel port to be in the ``on'' state, while all the other ports are in the ``off'' state.	
	\item {\textbf{All Off}}: This scheme sets all ports to be in the ``off'' state.
	\item {\textbf{All On}}: This scheme sets all ports to be in the ``on'' state.
\end{itemize}

\begin{figure}[t]
	\centering
	\includegraphics[width=8cm]{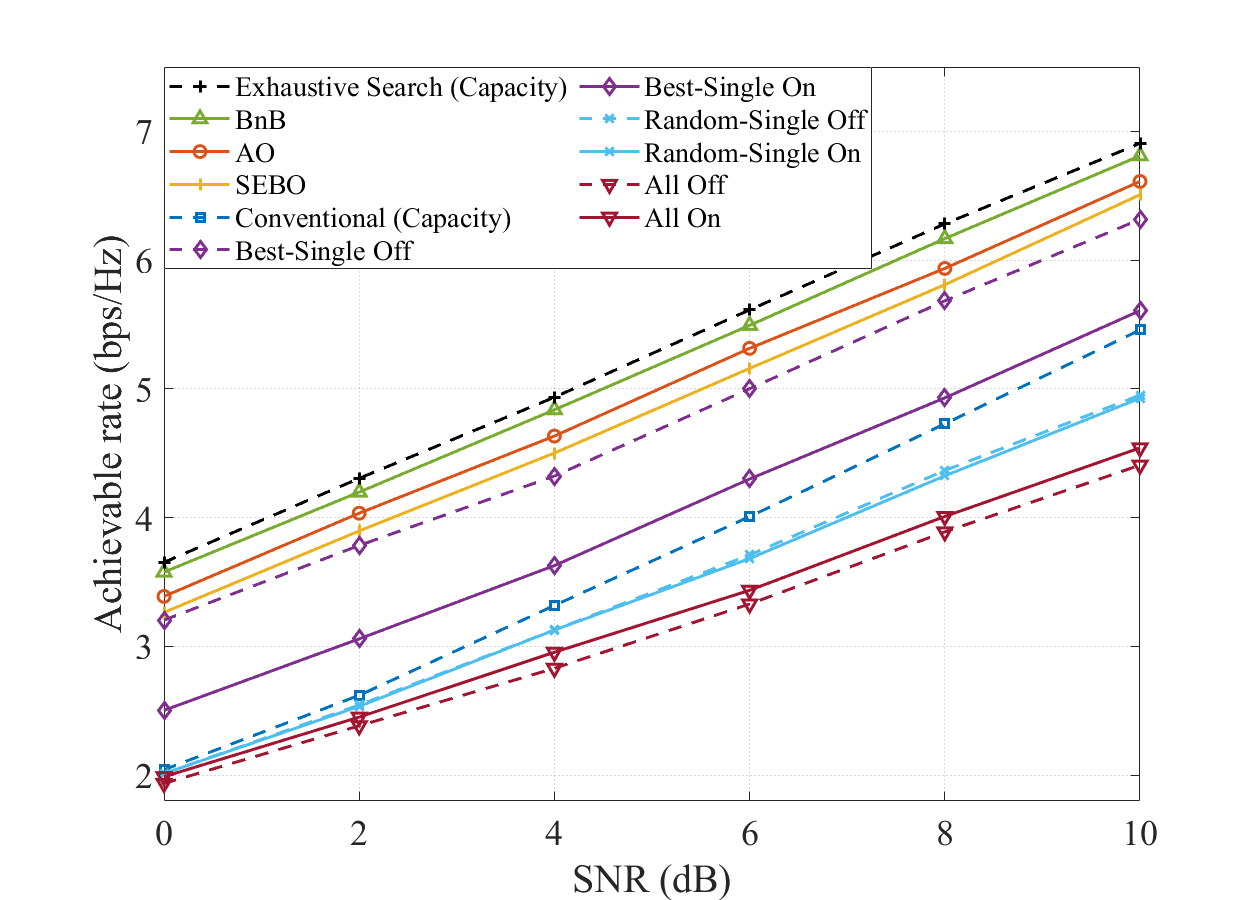}
	\vspace{-3mm}
	\caption{Achievable rate versus receive SNR.}
	\label{fig_SNR}
	\vspace{-8mm}
\end{figure}
Fig. \ref{fig_SNR} shows the achievable rate with the proposed schemes and benchmark schemes versus the receive SNR. It is observed that the proposed exhaustive search method achieves the highest achievable rate, since it corresponds to the fundamental capacity limit of pixel antenna based MIMO communication. The proposed branch-and-bound based algorithm performs closely to the exhaustive search method, which demonstrates its effectiveness. The element-wise AO based algorithm outperforms SEBO as well as all the other benchmark schemes, which indicates the high quality of its solution despite the polynomial complexity. On the other hand, it is observed that all the proposed schemes, SEBO, and the best-single off and on schemes outperform the capacity of conventional MIMO system without pixel antenna. This suggests that judicious optimization of the antenna coding can enhance the data rate due to the new DoF in radiation pattern reconfiguration. However, if the antenna coder is not properly designed (e.g., with random-single off/on or all off/on), pixel antenna based MIMO may be inferior to conventional MIMO, which further demonstrates the need of systematic optimization study on antenna coding. Finally, it is observed that the best-single off scheme and random-single off scheme outperform their ``on''-state counterparts. However, turning all the switches off will yield worse performance than turning them all on. This provides useful guidelines for practical heuristic antenna coding designs.

\begin{figure}[t]
	\centering
	\includegraphics[width=8cm]{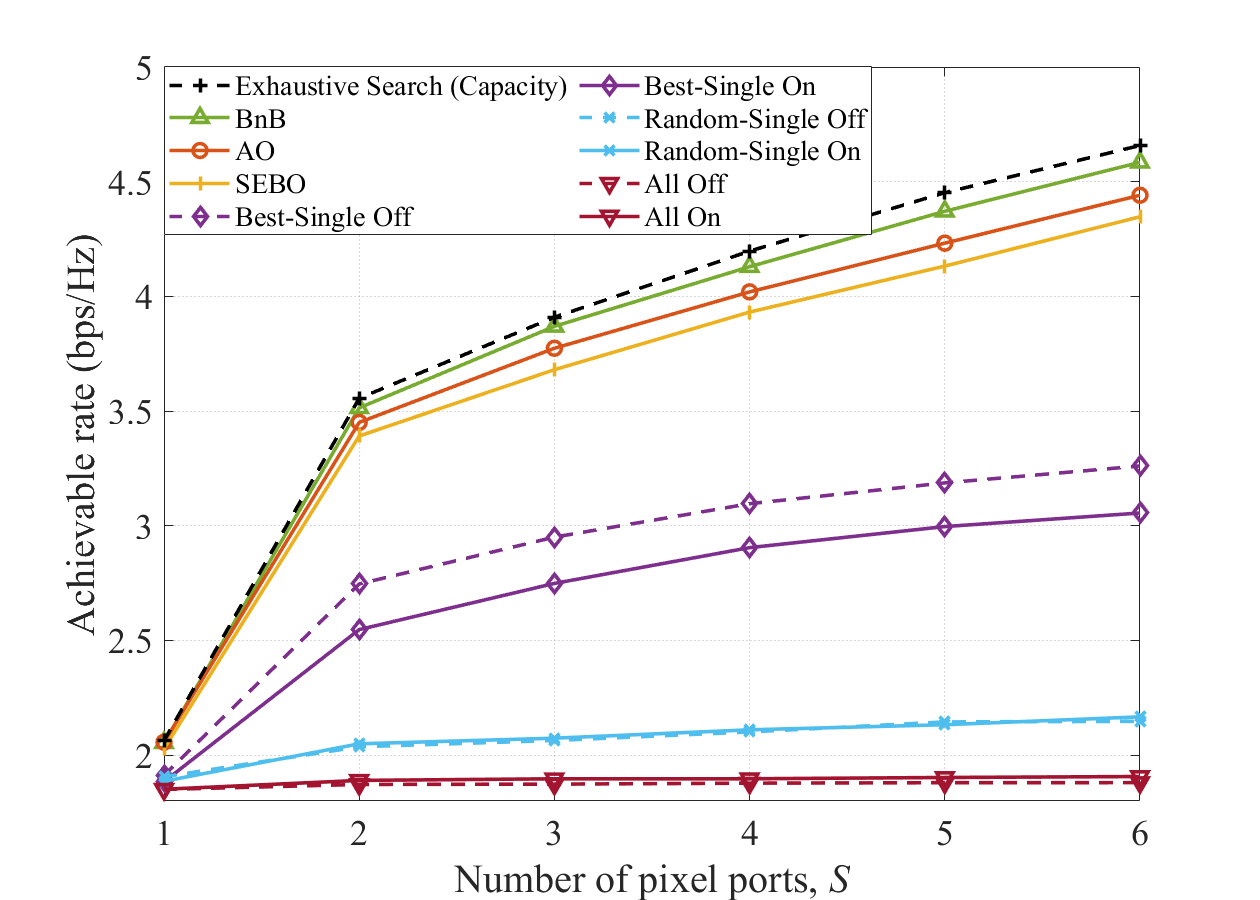}
				\vspace{-2mm}
	\caption{Achievable rate versus the number of pixel ports, $S$, at $\mathrm{SNR}=0$ dB.}
	\label{fig_S}
	\vspace{-6mm}
\end{figure}

\begin{figure}[t]
	\centering
	\includegraphics[width=8cm]{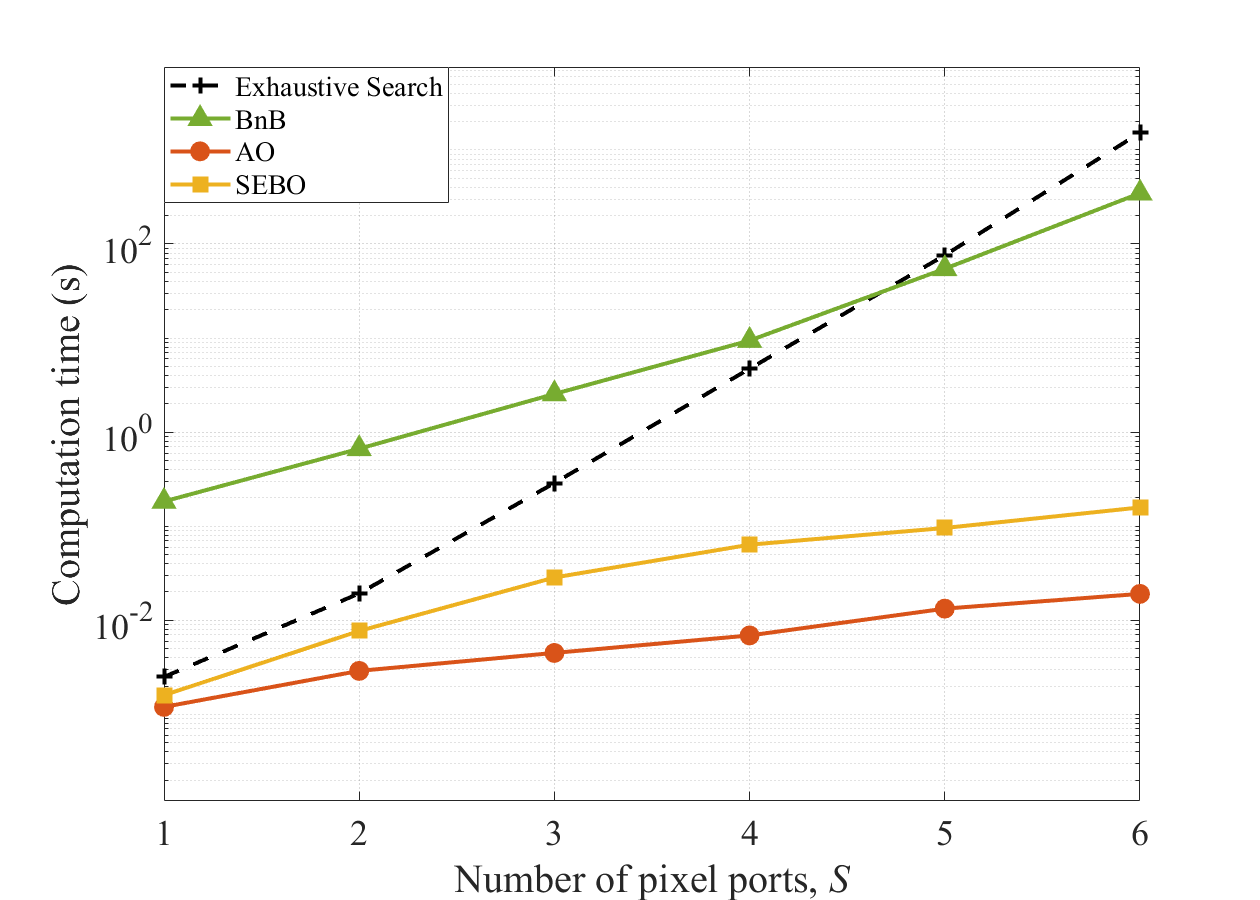}
				\vspace{-2mm}
	\caption{Computation time versus the number of pixel ports, $S$, at $\mathrm{SNR}=0$ dB.}
	\label{fig_time}
	\vspace{-7mm}
\end{figure}

Fig. \ref{fig_S} shows the achievable rate with the proposed schemes and benchmark schemes versus the number of pixel ports on each antenna, $S$, at $\mathrm{SNR}=0$ dB. Similar performance trends can be observed among different schemes. Particularly, as the number of pixel ports increases, the capacity gain of pixel antenna based MIMO communication over conventional MIMO communication ($S=0$) increases significantly. Moreover, the proposed branch-and-bound based algorithm, element-wise AO based algorithm, and SEBO also have more significant rate gains over conventional MIMO communication. Furthermore, as $S$ increases, the best-single off and best-single on schemes also perform drastically better. It is also observed that by simply increasing $S$ from $1$ to $6$, a dramatic rate gain of over $2.5$ bps/Hz can be achieved. This demonstrates the effectiveness of pixel antennas in enhancing the data rate even with a small number of pixel ports or switches.

Finally, Fig. \ref{fig_time} shows the average computation time for the proposed schemes and SEBO versus the number of pixel ports on each antenna, $S$, at $\mathrm{SNR}=0$ dB. It is observed that as $S$ increases, the exhaustive search method consumes the most computation time, and the element-wise AO based algorithm is the most time efficient, which are consistent with our complexity analysis. On the other hand, the branch-and-bound based algorithm consumes more time for very small values of $S$, due to the need to iteratively solve (P1-B) and (P1-Q) before convergence. To summarize, the proposed exhaustive search, branch-and-bound, and element-wise AO based algorithms can achieve a flexible trade-off between performance and complexity. 

\section{Conclusions}
This paper studied the capacity characterization problem for pixel antenna based MIMO communication, via the joint optimization of the binary transmit and receive antenna coders and the transmit covariance matrix. Despite the challenging MINLP nature and non-convexity of the problem, both the optimal solution via exhaustive search and two suboptimal solutions via branch-and-bound and element-wise AO have been proposed, which were shown to achieve a flexible trade-off between performance and complexity. Numerical results showed that the proposed algorithms outperform various benchmark schemes, and pixel antenna based MIMO communication outperforms conventional MIMO communication in terms of capacity with properly designed antenna coders.

\end{document}